\documentclass{aastex631}
\usepackage{graphicx}
\usepackage{natbib}

\begin{document}

\title{Beyond BPT: A New Multi-Dimensional Diagnostic Diagram for Classifying Power Sources Tested Using the SAMI Galaxy Survey}

\author{Victor D. Johnston}
\affiliation{Ritter Astrophysical Research Center and Department of Physics \& Astronomy, University of Toledo, Toledo, OH 43606, USA}

\author[0000-0001-7421-2944]{Anne M. Medling}
\altaffiliation{Hubble Fellow}
\affil{Ritter Astrophysical Research Center and Department of Physics \& Astronomy, University of Toledo, Toledo, OH 43606, USA}
\affil{ARC Centre of Excellence for All Sky Astrophysics in 3 Dimensions (ASTRO 3D)}

\author[0000-0002-9768-0246]{Brent Groves}
\affil{International Centre for Radio Astronomy Research, The University of Western Australia, 35 Stirling Highway, Crawley WA 6009, Australia}

\author[0000-0001-8152-3943]{Lisa J. Kewley}
\affil{Center for Astrophysics — Harvard \& Smithsonian, 60 Garden Street, Cambridge, MA 02138, USA}
\affil{ARC Centre of Excellence for All Sky Astrophysics in 3 Dimensions (ASTRO 3D)}

\author[0000-0002-7422-9823]{Luca Cortese}
\affil{International Centre for Radio Astronomy Research, The University of Western Australia, 35 Stirling Highway, Crawley WA 6009, Australia}
\affil{ARC Centre of Excellence for All Sky Astrophysics in 3 Dimensions (ASTRO 3D)}

\author[0000-0003-2880-9197]{Scott Croom}
\affil{Sydney Institute for Astronomy, School of Physics, A28, The University of Sydney, NSW 2006, Australia}
\affil{ARC Centre of Excellence for All Sky Astrophysics in 3 Dimensions (ASTRO 3D)}

\author[0000-0001-8083-8046]{Ángel R. López-Sánchez}
\affil{Department of Physics and Astronomy, Macquarie University, NSW 2109, Australia}
\affil{School of Mathematics and Physical Sciences, Macquarie University, NSW 2109, Australia}
\affil{Macquarie University Research Centre for Astronomy, Astrophysics \& Astrophotonics, Sydney, NSW 2109, Australia}
\affil{ARC Centre of Excellence for All Sky Astrophysics in 3 Dimensions (ASTRO 3D)}

\author[0000-0003-4334-9811]{Henry Zovaro}
\affil{Research School of Astronomy \& Astrophysics, Mount Stromlo Observatory, Australia National University, Cotter Road, Weston, ACT 2611, Australia}
\affil{ARC Centre of Excellence for All Sky Astrophysics in 3 Dimensions (ASTRO 3D)}

\author[0000-0001-7516-4016]{Joss Bland-Hawthorn}
\affil{Sydney Institute for Astronomy, School of Physics, A28, The University of Sydney, NSW 2006, Australia}

\author[0000-0003-1627-9301]{Julia Bryant}
\affil{Sydney Institute for Astronomy, School of Physics, A28, The University of Sydney, NSW 2006, Australia}

\author[0000-0002-6998-6993]{Jon Lawrence}
\affil{Department of Physics and Astronomy, Macquarie University, NSW 2109, Australia}

\author[0000-0002-2879-1663]{Matt Owers}
\affil{Department of Physics and Astronomy, Macquarie University, NSW 2109, Australia}

\author[0000-0002-5368-0068]{Samuel Richards}
\affil{Sydney Institute for Astronomy, School of Physics, A28, The University of Sydney, NSW 2006, Australia}

\author[0000-0003-2552-0021]{Jesse van de Sande}
\affil{Sydney Institute for Astronomy, School of Physics, A28, The University of Sydney, NSW 2006, Australia}
\affil{ARC Centre of Excellence for All Sky Astrophysics in 3 Dimensions (ASTRO 3D)}

\setlength{\parskip}{1em}

\begin{abstract}

Current methods of identifying the ionizing source of nebular emission in galaxies are well defined for the era of single fiber spectroscopy, but still struggle to differentiate the complex and overlapping ionization sources in some galaxies. With the advent of integral field spectroscopy, the limits of these previous classification schemes are more apparent. We propose a new method for distinguishing the ionizing source in resolved galaxy spectra by use of a multi-dimensional diagnostic diagram that compares emission line ratios with velocity dispersion on a spaxel by spaxel basis within a galaxy. This new method is tested using the SAMI Galaxy Survey Data Release 3, which contains 3068 galaxies at z $<$ 0.12. Our results are released as ionization maps available alongside the SAMI DR3 public data. Our method accounts for a more diverse range of ionization sources than the standard suite of emission line diagnostics; we find 1433 galaxies with significant contribution from non-star-forming ionization using our improved method as compared to 316 galaxies identified using only emission line ratio diagnostics. Within these galaxies, we further identify 886 galaxies hosting unique signatures inconsistent with standard ionization by \ion{H}{2} regions, AGN, or shocks. These galaxies span a wide range of masses and morphological types and comprise a sizable portion of the galaxies used in our sample. With our revised method, we show that emission line diagnostics alone do not adequately differentiate the multiple ways to ionize gas within a galaxy.

\end{abstract}

\section{Introduction}

Investigating the prevalence and effects of different ionizing sources within galaxies is a topic that has been of interest for decades \citep[see][for a review]{Kewley2019}. Separating different ionization mechanisms from one another in large extra-galactic surveys has been primarily by use of optical diagnostic diagrams. These diagnostics compare key spectral features that are largely distinct for each ionizing source. For example, regions of high star formation ionize the surrounding \ion{H}{2} regions with the radiation released from O and B stars. Active galactic nuclei (AGN) emit a hard X-ray radiation field that is also capable of exciting nearby gas, to higher energy levels than star formation can. Supernovae and galactic winds can shock-heat gas, collisionally ionizing it.

The most prevalent optical emission line diagnostic diagrams are the BPT \citep*{1981PASP...93....5B} and VO87 \citep{1987ApJS...63..295V} diagrams, which compare the flux ratios of emission line pairs log([\ion{O}{3}]~$\lambda$5007/H$\beta$) and log([\ion{O}{1}]~$\lambda$6300/H$\alpha$), log([\ion{N}{2}]~$\lambda$6583/H$\alpha$), and log([\ion{S}{2}]~$\lambda\lambda$6716,31/H$\alpha$), and the WHAN diagram \citep{WHAN} that compares the log([\ion{N}{2}]~$\lambda$6583/H$\alpha$) ratio to the equivalent with of H$\alpha$. These diagrams have proven useful as the different nature of each ionization source leads to different physical conditions in the ionized gas. The hard radiation field from an AGN produces higher ionization states in the surrounding interstellar medium as well as higher temperatures, increasing the strength of collisionally excited lines such as [\ion{O}{3}] relative to the recombination lines like H$\beta$. Shocks have similarly elevated emission in the forbidden lines due to high temperatures, but also have extended zones of hot, partially ionized gas that lead to different flux ratios than AGN \citep{1987ApJS...63..295V}.

These classification schemes were refined and popularized using single fiber nuclear spectra from the Sloan Digital Sky Survey \citep[SDSS;][]{SDSSDR4}, showing that the central excitation source varies across the population of local galaxies. Star formation-dominated galaxies have a metallicity-dependent track that lies below the maximum theoretical starburst line, while AGN or shock-dominated galaxies are shifted upward due to the increased ionization parameter \citep{2001ApJ...556..121K,2003MNRAS.346.1055K,2006MNRAS.371.1559G,2006MNRAS.372..961K,MappingsIII_2008,2015_Medling,2015ApJS..221...28R,Sanchez_2015,Kewley2019}.

In addition to the use of emission line diagnostics, kinematic information can be used to help identify ionization sources within a galaxy. It is well documented that shocks, due to the turbulent nature of collisions, have an increased velocity dispersion that correlates to the strength of the shock as compared to the relatively lower velocity dispersion of nearby star-formation dominated regions \citep{2006_Monreal,2010_Monreal,2011_Rich,2014_Rich,2021_Law}. These mechanisms have proved useful in identifying gas entrained within and ionized by outflows \citep{2019_Lopez,2020_Lopez,2021_Law,2022_Law}. Location within a galaxy, such as galactocentric radius, provides another useful check for providing information on ionization sources. AGN ionization tends to occur closer to the nuclear region of a galaxy while shocks are often associated with outflows and galaxy collisions \citep{2010_Sharp,2018_D'Agostino,2018_Sanchez,2020_Lacerda}.

Integral field spectroscopy (IFS) has enabled studies that compare these physical mechanisms across different spatial locations in a single galaxy \citep[e.g.][]{2014MNRAS.444.3894H,2015ApJS..221...28R,2019MNRAS.486..344R,Rich2011,Rich2014,Husemann2014,Sharp_2010,Rebecca2015,Davies2014A,Davies2014B}. In order to gain a better understanding of the physical mechanisms that drive galaxy evolution, we require detailed information about not only what is happening inside of them, but where as well.

In the past two decades, different IFS surveys have provided a wealth of simultaneous spatial and spectral information about galaxies. The Calar Alto Legacy Integral Field Area Survey (CALIFA) has nearly 700 galaxy observations, with a subset focused on rarer galaxy types such as dwarf galaxies and low and high mass early type galaxies \citep{CALIFA}. The Mapping Nearby Galaxies at APO Survey (MaNGA) is an extension of SDSS that has observed roughly 10,000 galaxies of all sizes and types, \citep{Manga,SDSS_DR17}. The Sydney-Australian-Astronomical-Observatory Multi-object Integral-Field Spectrograph Galaxy Survey (SAMI) contains over 3000 galaxies with a wide range of masses and morphological types in the nearby universe \citep{SAMIDR3}. These surveys, and others of their kind, provide the required depth of data to investigate ionizing sources on large scales that were previously impossible.

A recent example of the power of IFS data, in particular its ability to help identify ionization sources, can be seen in the work of \citet{2019MNRAS.487.4153D}. Created using data of NGC\,1068 taken from the Siding Spring Southern Seyfert Spectroscopic Snapshot Survey \citep{S7DR2}, these authors shows how using spatially resolved emission line fluxes, as well as derived gas kinematics and physical radius can be used to separate regions of pure star-formation, AGN, and shock ionization from one another. This technique uses a combination of several methods discussed above in order to reduce the amount of degeneracy that any one individual method has in determining ionization sources. Indeed these basic principles form the basis of our work presented below.

In this paper, we demonstrate a new automated classification method that uses a combination of emission lines and kinematics to separate star formation, AGN, and shocks on a spatially resolved scale. We also identify a number of populations that are distinct from the standard interpretation of star-forming galaxies being dominated by young stellar emission. Section \ref{SAMI} describes the data used from Data Release 3 of the SAMI Galaxy Survey. Section \ref{Breaking shock} reviews the limitations of existing emission line diagnostic methods. Section \ref{New_Scheme} contains the methods and implementations of the new classification scheme. Section \ref{compare} compares the results of our new schema to the classical emission line diagnostics and explains the advantages of using additional kinematic and spatial information. Finally, Section \ref{ionization_maps} contains an overview of the shock map data products that we are releasing and Section \ref{conclusion} is our conclusions.

\section{The SAMI Galaxy Survey} \label{SAMI}

We use optical integral field spectroscopy collected from the Sydney-Australian-Astronomical-Observatory Multi-object Integral-Field Spectrograph (SAMI) on the 3.9-meter Anglo-Australian Telescope at Siding Spring Observatory \citep{2012MNRAS.421..872C}. The SAMI instrument observes each target with a fiber bundle consisting of 61 fibers allowing for a field of view (FOV) of 15 arcsec per observation \citep{Hexabundles,Hexabundles2}. The SAMI Galaxy Survey Data Release 3 contains 3068 galaxies \citep{SAMIDR3} and spans a wide mass range from log(M\textsubscript{$\star$}/M\textsubscript{$\odot$}) of 7.5 to 11.6 and a range in redshift of 0.004 $<$ z $<$ 0.113 \citep{2018MNRAS.481.2299S}. The galaxies observed with SAMI were chosen from the Galaxy And Mass Assembly (GAMA) Survey \citep{GAMA}. The SAMI instrument is composed of a blue and red arm that span the majority of the visible wavelength range. For the SAMI Galaxy survey we use the 580V grating at 3750-5750$\AA$ giving a resolution of R = 1808 ($\sigma$=70.4 km s$^{-1}$), and the R1000 grating from 6300-7400A giving a resolution of R = 4304 ($\sigma$=29.6 km s$^{-1}$) \citep{van_de_Sande_2017b,2018MNRAS.481.2299S}. The resulting spatial scale of the data cube is 0{\farcs}5 per pixel with a 50 by 50 image for a total FOV of 25{\arcsec} x 25{\arcsec}. Galaxies from a wide range of environments were collected, from isolated field galaxies to large groups and clusters \citep{SAMISelection,SAMIClusters}.

The emission line fitting was performed using the LZIFU software package \citep{2016Ap&SS.361..280H,2018MNRAS.475.5194M}. Each spectrum is fit with the following 11 strong lines simultaneously; H$\alpha$, H$\beta$, [\ion{O}{3}]$~\lambda\lambda$4959, 5007, [\ion{O}{1}]~$\lambda$6300, [\ion{N}{2}]~$\lambda\lambda$6548, 83, [\ion{S}{2}]~$\lambda\lambda$6716, 31, and [\ion{O}{2}]~$\lambda\lambda$3726, 29, with up to three Gaussian components for each, and merged using the trained neural net LZCOMP \citep{2017MNRAS.470.3395H}. All ionized gas species are constrained to have the same kinematic profiles. Multiple components are fit in order to account for the kinematically distinct phases of the ionized gas, such as outflowing material, allowing for a more accurate representation for the gas emission. The resulting data products are two dimensional spatial maps where emission line flux, velocity, and velocity dispersion are extracted. For our diagnostics we adopt a signal to noise minimum of 3 for each emission line used in our analysis. We also only include galaxies that have 10 or more spaxels remaining after this cut. This cut ensures that the results of our analysis are not skewed by faint, low signal galaxies. We also calculate a projected galactocentric radius of each spaxel by using the redshift and inclination data provided in the SAMI catalogs for each galaxy \citep{2017MNRAS.468.1824O,SAMISelection}.

\section{Breaking the Shock Degeneracies} \label{Breaking shock}

The spatially-resolved nature of IFS has proven useful in identifying trends relating to the ionization structure of galaxies. Excitation from AGN tend to be found toward the central region of a galaxy, though it is possible that the hard ionizing field can dominate further out into the galaxy \citep{2014ApJ...787...65H}. Shocked gas, however, can be linked to many different physical processes and is therefore not expected to be contained in any particular region of a galaxy. When multiple excitation sources are present within a galaxy, their spatially resolved emission line diagnostic diagrams produce mixing sequences, showing the progression from the star-forming region into the AGN and LINER regions of the emission line diagnostics as the relative fraction of ionization from AGN or shocks increases \citep{Davies2014A,Davies2014B}.

Shocked gas is not well separated by the BPT/VO87 diagnostic diagrams, because its line ratios are degenerate with those of AGN \citep{2008ApJS..178...20A,2016ApJ...833..266T}. Because of this overlap in emission line space, it is impossible to separate them without additional information. One such way to differentiate shocks from other excitation sources is to look at the kinematics to see how turbulent regions of a galaxy are. Given that shock-excitation is a mechanical process rather than radiative, shocked gas tends to be more kinematically disturbed. It has been shown that gas ionized by stars (or gas within \ion{H}{2} regions) has significantly lower velocity dispersions, typically less than 50 km\,s$^{-1}$ as compared to shocked gas which has 100 km\,s$^{-1}$ or higher \citep{Rich2011,2015ApJS..221...28R,2014MNRAS.444.3961D}.

An additional challenge in characterizing gas excitation sources is that it is difficult to examine a large number of galaxies at once. The combination of line ratios, kinematics, and spatial information has so far proved to be the most reliable method of determining the power sources of galactic outflows. A way to quickly classify outflows while keeping the multiple dimensions of information was proposed by \citet{2019MNRAS.487.4153D} in a new 3D diagnostic diagram. This diagram uses velocity dispersion, projected radius, and a modified line ratio to determine two distinct tracks from star formation to shock and AGN excitation. Tested on an actively star-forming galaxy that has a known AGN and shows shock activity, this method provides a great example of the new level of analysis that is possible with IFS. Because shocks and AGN lie in separate regions in this diagram, it is possible to determine the separate mixing sequences of both. This allows for a fractional contribution of each power source to be determined at each spatial location. The degeneracies that were present on the BPT/VO87 diagrams are resolved here. 

Though this method easily breaks the degeneracies in the extreme case of NGC\,1068, it does not separate AGN-ionized and shocked-ionized gas in the more moderate SAMI sample. This degeneracy is likely due to several reasons, the most prominent being that due to the close distance of NGC\,1068, the spatial resolution of the data ($\sim$100 pc/px) is at a much smaller physical scale then that of a typical SAMI galaxy (median of 1.65 kpc/px). This drastic improvement in spatial resolution allows better separation between the multiple sources of ionization found within a galaxy that would otherwise be blended together on larger scales. This blending prevents their classification scheme from being broadly applicable to data from surveys with spatial resolutions worse than $\sim$1 kpc/px. We are therefore motivated to extend the 3D diagnostic diagram technique to find one sensitive enough to classify SAMI galaxies, and other current large IFS surveys, that works on both small and moderate spatial scales.

\section{New 3D Classification scheme} \label{New_Scheme}

We propose a new classification method using a combination of emission line ratios and kinematics for determining the main excitation source of warm ionized gas. Our new diagnostic uses the same emission line ratios as the standard BPT/VO87 diagrams, while adding a third axis of velocity dispersion to further investigate the physical nature of the gas. By using this combination of parameters, we are able to detect fainter signatures of shocks and AGN than with any other method. We also detect multiple populations of galaxies whose classifications from their emission line ratios are inconsistent with the physical interpretation given by their velocity dispersion. We do this by identifying and tracing the mixing sequence from star-forming gas to either a shock track or an AGN track. By following these mixing sequences we can extract the relative shock or AGN fraction for an individual spaxel within a galaxy on a per galaxy basis. 

The first step in our new method involves identifying galaxies that have signatures of AGN-like ionization in all 3 of the BPT/VO87 diagrams. We use a criteria of requiring 10 spaxels in each of the AGN regions and do not use any velocity dispersion cutoffs for this group. This is to ensure that galaxies that are being ionized primarily by the hard radiation field of the black hole with no shock component are not missclassified due to lacking any elevated velocity dispersion. We use the equivalent width of the emission of H$\alpha$ as an additional parameter to further separate AGN ionization from contamination from regions of hot low mass evolved stars (HOLMES). If 70\% of the AGN-like spaxels have an equivalent width of 3 or less, the galaxy is classified as being dominated by HOLMES emission, otherwise it is considered a true AGN.

We next examine the distribution of velocity dispersions for signatures of multiple kinematic components in each galaxy individually. For this we use the multiple component spectral fits provided by LZIFU. We then use a multi-component Gaussian fit to the histogram of the velocity dispersion to see if there is a population of gas that is distinct from the standard, low velocity dispersion star-forming regions. If the range of velocity dispersion is best fit with two components, or if the peak of the single component minus one standard deviation is greater than 50 km s$^{-1}$, the gas is flagged as kinematically disturbed. Both the kinematically disturbed and non-kinematically disturbed gas are then examined using their emission line ratios.

Starting with the kinematically disturbed galaxies, we see where the spaxels lie on the classical BPT/VO87 diagrams. To be considered as containing significant contribution from an ionization source, we require 10 spaxels or more in a given region for 2 out of 3 of the log([\ion{N}{2}]/H$\alpha$), log([\ion{S}{2}]/H$\alpha$) and log([\ion{O}{1}]/H$\alpha$) diagnostics. This leads to 4 cases; AGN-like, LINER-like, AGN- and LINER-like, and star-forming-like ionization. For AGN-like ionization we again check the equivalent width of H$\alpha$ to seperate out HOLMES contamination. If no significant HOLMES contribution is found the galaxy is classified as an AGN, otherwise it is moved to the HOLMES category. We use this 2 out of 3 emission line plus velocity dispersion criteria for these AGN as these galaxies are likely ones that host AGN with mild outflows. For the LINER-like galaxies, we classify all of these as being shock ionized as the emission line ratios and their high velocity dispersion match the classic case of turbulent, collisionally excited gas. Galaxies with AGN- and LINER-like ionization are checked once again for HOLMES contamination, and are labeled as shock ionized if the AGN-like spaxels are contaminated with HOLMES, and as AGN if they are not. For these galaxies with AGN- and LINER-like ionization not contaminated by HOLMES, we label them as AGN even though they show significant signatures of shock excitation as they are most likely galaxies that host strong outflows from the central AGN region. Finally, for galaxies that do not contain significant signatures of non-star-forming like ionization, we label these as their own category of elevated velocity dispersion, low emission line ratio. As their velocity dispersions are not consistent with pure \ion{H}{2} ionization, we are hesitant to classify all of these galaxies as simply star-forming. A further discussion of this group is located in section \ref{ElKiLoEm}.

Now we look at the galaxies with no elevated velocity dispersion. Again, we use a criteria of 2 out of 3 regions of the emission line diagnostics as explained above. This presents 3 cases instead of 4 this time; AGN-like, LINER-like, and star-forming-like. for AGN-like galaxies we perform the same HOLMES check as in the other AGN-like cases and label those without HOLMES contamination as simply hosting AGN-like ionization. It is worth explicitly stating that we do not label these galaxies as containing true AGN, but only as hosting AGN-like ionization. Our reasoning for this is explained in section \ref{AGN_like}. For LINER-like galaxies, we classify them as hosting ionization from diffuse ionized gas as their low velocity dispersions are inconsistent with shock ionized gas. The remainder of the galaxies are classified as classical star-forming galaxies as they have both low velocity dispersions and emission line ratios.

\begin{figure}[ht]
 \centering
   \includegraphics[width=.95\linewidth]{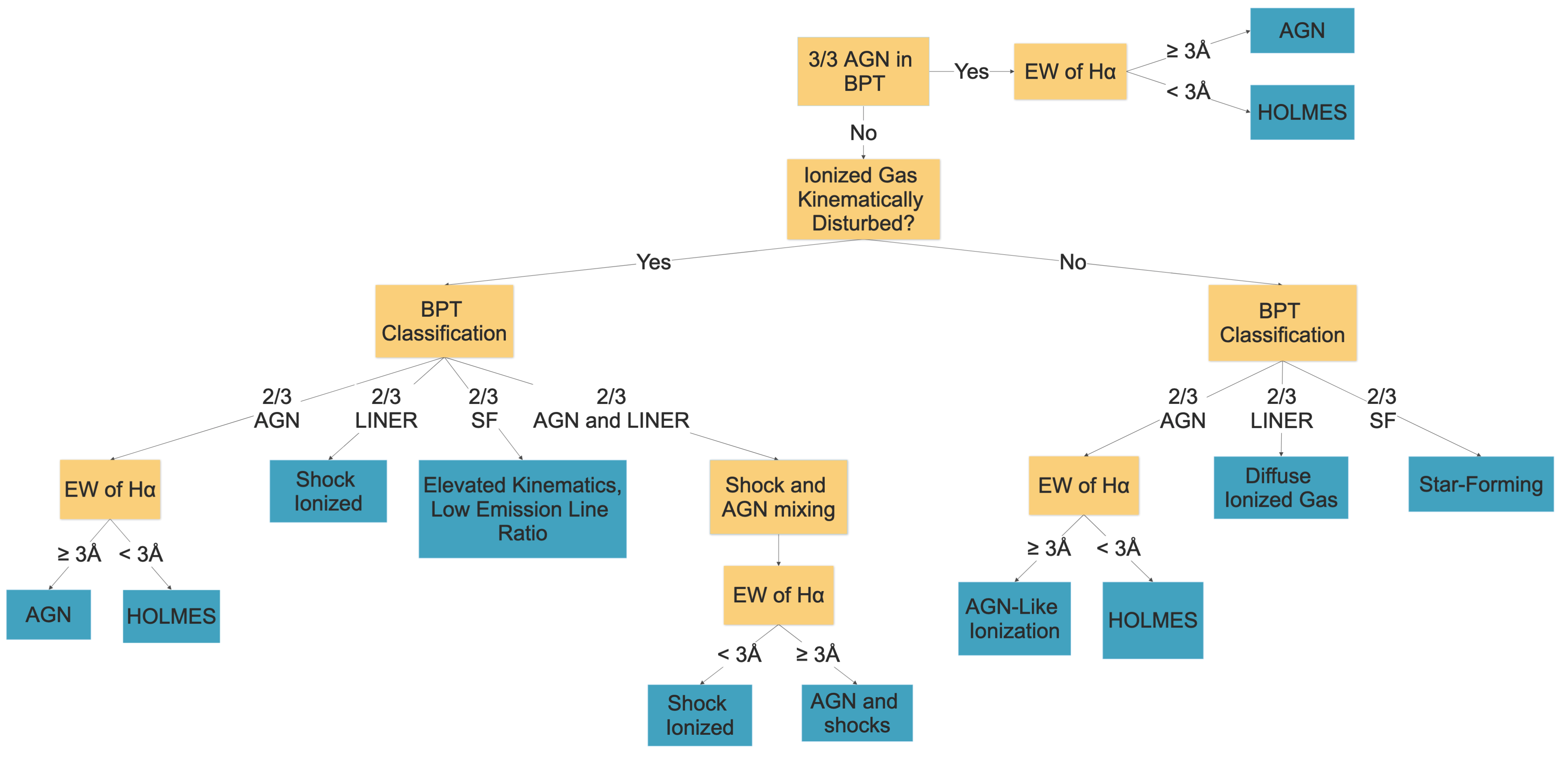}
   \caption{A flowchart showing a simplified decision tree used by our classification method. The yellow boxes show steps along the classification process while the other colored boxes represents the final classification that a galaxy can fall into.} 
   \label{flowchart}
 \end{figure}

A flowchart showing an overview of the classification scheme is shown in Figure \ref{flowchart} for quick reference and an in depth breakdown of each category can be found in section \ref{compare}.

\begin{figure}[ht]
 \centering
   \includegraphics[width=.95\linewidth]{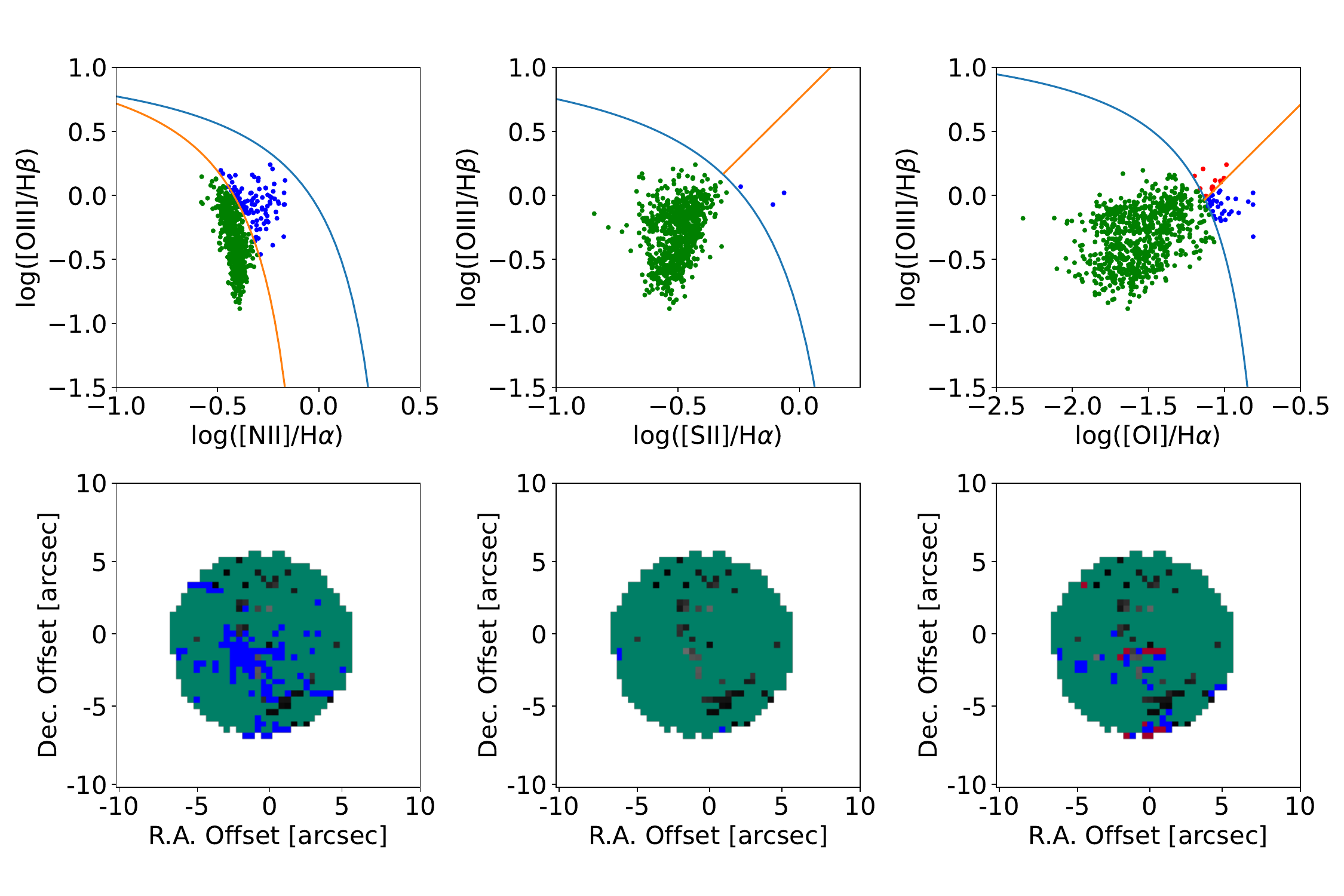}
   \caption{Emission line diagnostics and spatial maps for GAMA 106717. Each figure in the first row shows the emission line diagnostics colored according to where each spaxel falls in their respective diagram, green for star-forming, blue for composite or LINER, and red for AGN. Below each diagnostic is the corresponding spatial map showing where in the galaxy the points in the diagram above are located.} 
   \label{106717_2d}
 \end{figure}

\begin{figure}[ht]
 \centering
   \includegraphics[width=.95\linewidth]{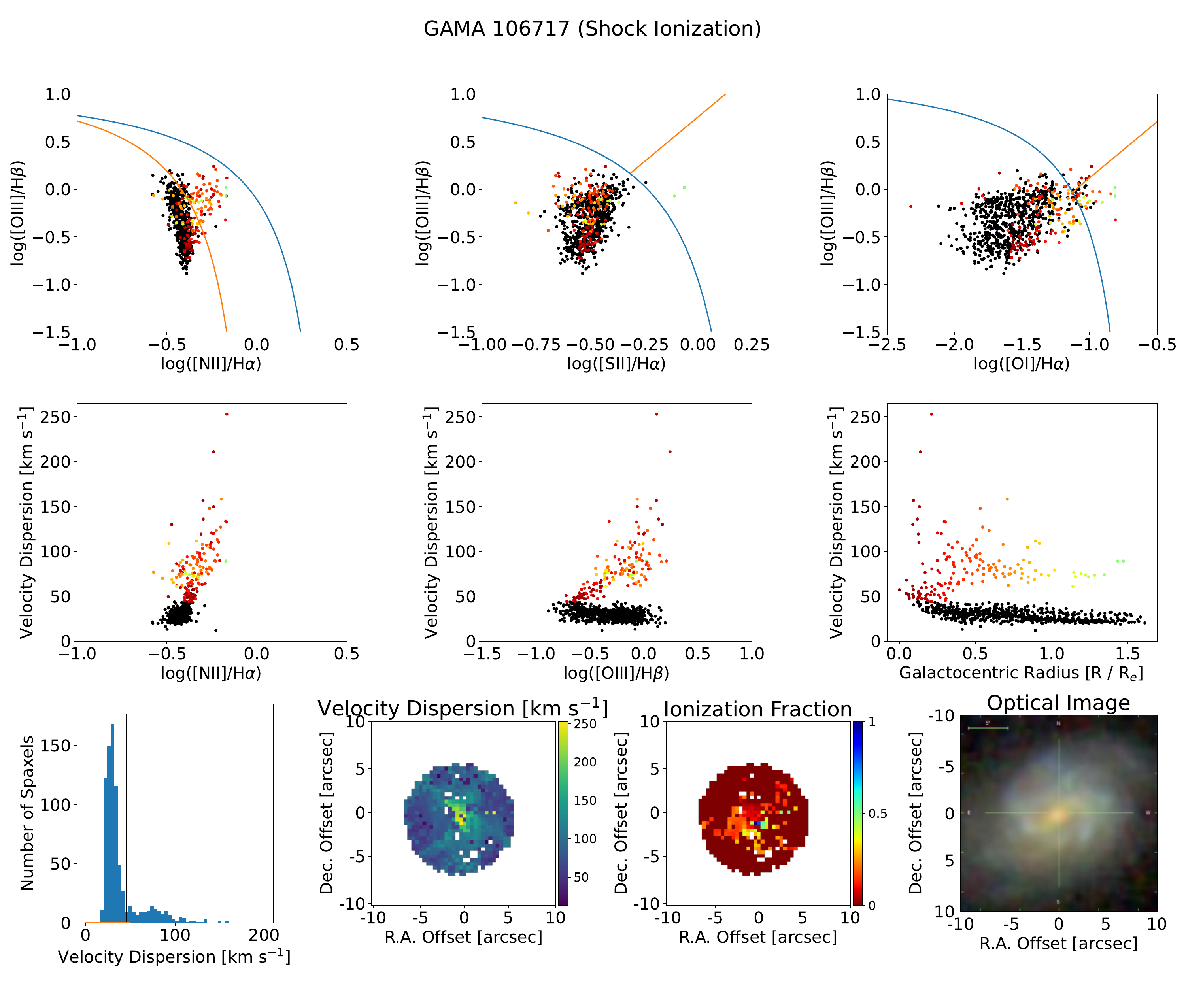}
   \caption{Emission line diagnostics and maps for GAMA 106717. The first row are the BPT/VO87 diagrams colored according to our new diagnostic, with black points being spaxels dominated by star-formation and colored points are spaxels with shock ionization colored by increasing radius from the center of the galaxy. The middle row contains plots of log([\ion{N}{2}]/H$\alpha$), log([\ion{O}{3}]/H$\beta$), and effective radius versus velocity dispersion. The separation between shocked and star-forming gas can be easily seen in the middle and right figure in this row. The last row contains a histogram of the velocity dispersion with a vertical line separating the low and high groups of gas, a map of the velocity dispersion, A shock fraction map which shows the relative contribution of shocks on a per spaxel basis, and a three color image on the right matching the FOV of the SAMI data.} 
   \label{106717_panels}
 \end{figure}

We describe two examples of a galaxy hosing either shock or AGN ionization here, for GAMA galaxies 106717 and 376478 respectively. Figure \ref{106717_2d} shows the BPT/VO87 diagnostics as well as a spatial map for each diagram for GAMA 106717, which contains regions of shocked gas. It is clear that this galaxy hosts LINER ionization that is spatially consistent between multiple emission line ratio pairs. Figure~\ref{106717_panels} shows the results of our new classification scheme. The first row is the BPT/VO87 diagrams with spaxels labeled according to our new method, with black points having only star-forming ionization and the colored points are the high velocity dispersion spaxels colored according to increasing radius, which for this galaxy contains shock ionization. The middle row is velocity disperision versus log([\ion{N}{2}]/H$\alpha$), log([\ion{O}{3}]/H$\beta$), and projected galactocentric radius colored the same as the plots above. From the projections in velocity dispersion space, two distinct populations of gas can be seen that start at low velocity dispersion and increase with emission line ratios. Two mixing sequences can be drawn from these plots, one for the low velocity disperion star-forming gas, and another for the high velocity dispersion shocked gas that both begin at a single point at the base of the star-forming sequence in the log([\ion{N}{2}]/H$\alpha$) diagram. The last row contains a histogram and a map of the velocity dispersion, a map of the shock fraction which is described in more detail in Section~\ref{ionization_maps}, and an optical image of the galaxy for reference. The combination of kinematic information as well as emission lines are what allow us to detect fainter signatures of shock ionization in this galaxy that would be missed with emission line ratios alone.

\begin{figure}[ht]
 \centering
   \includegraphics[width=.95\linewidth]{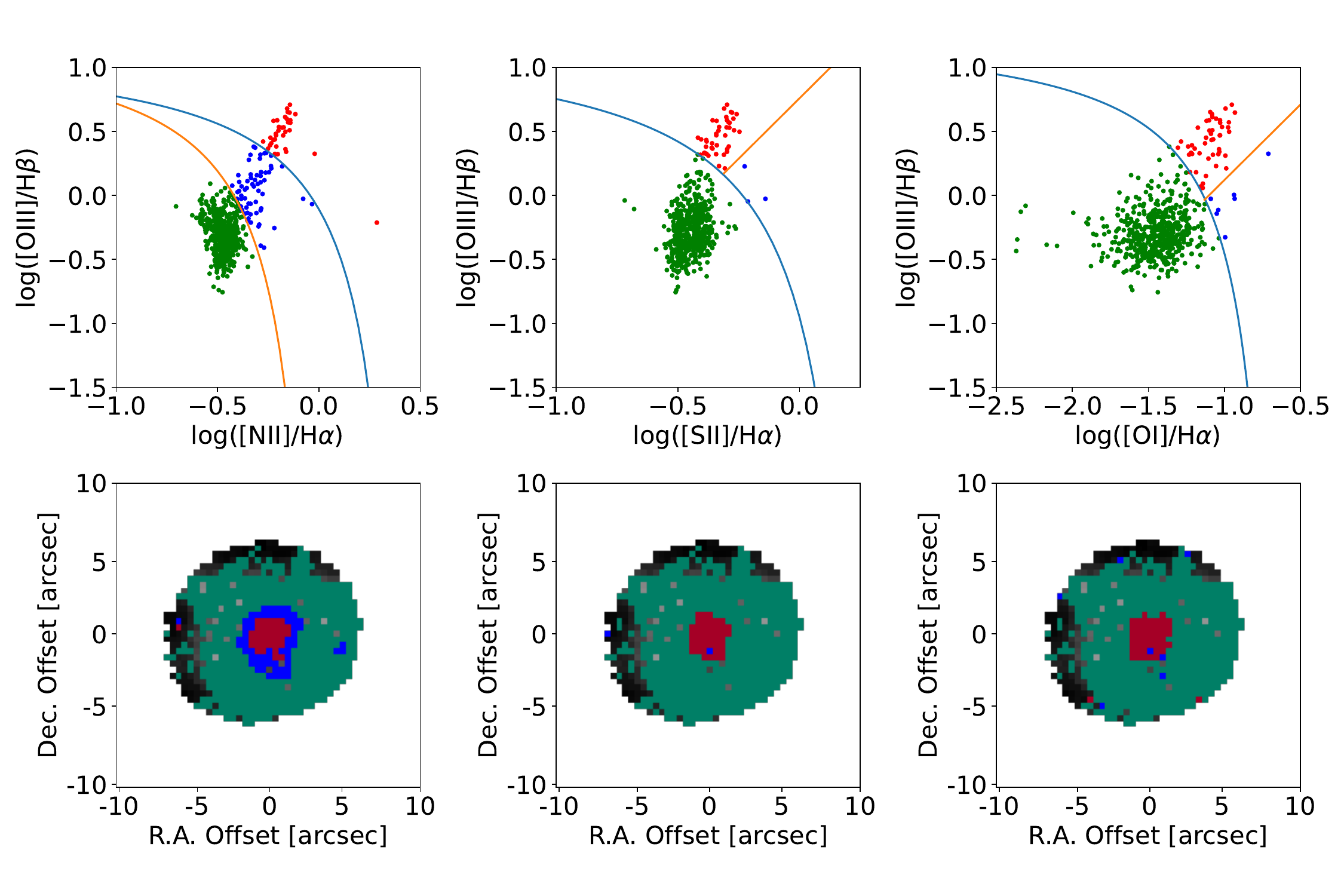}
   \caption{Emission line diagnostics and spatial maps for GAMA 376478. Each figure in the first row shows the emission line diagnostics colored according to where each spaxel fall in their respective diagram, green for star-forming, blue for composite or LINER, and red for AGN. Below each diagnostic is the corresponding spatial map showing where in the galaxy the points in the diagram above are located.} 
   \label{376478_2d}
 \end{figure}

\begin{figure}[ht]
 \centering
   \includegraphics[width=.95\linewidth]{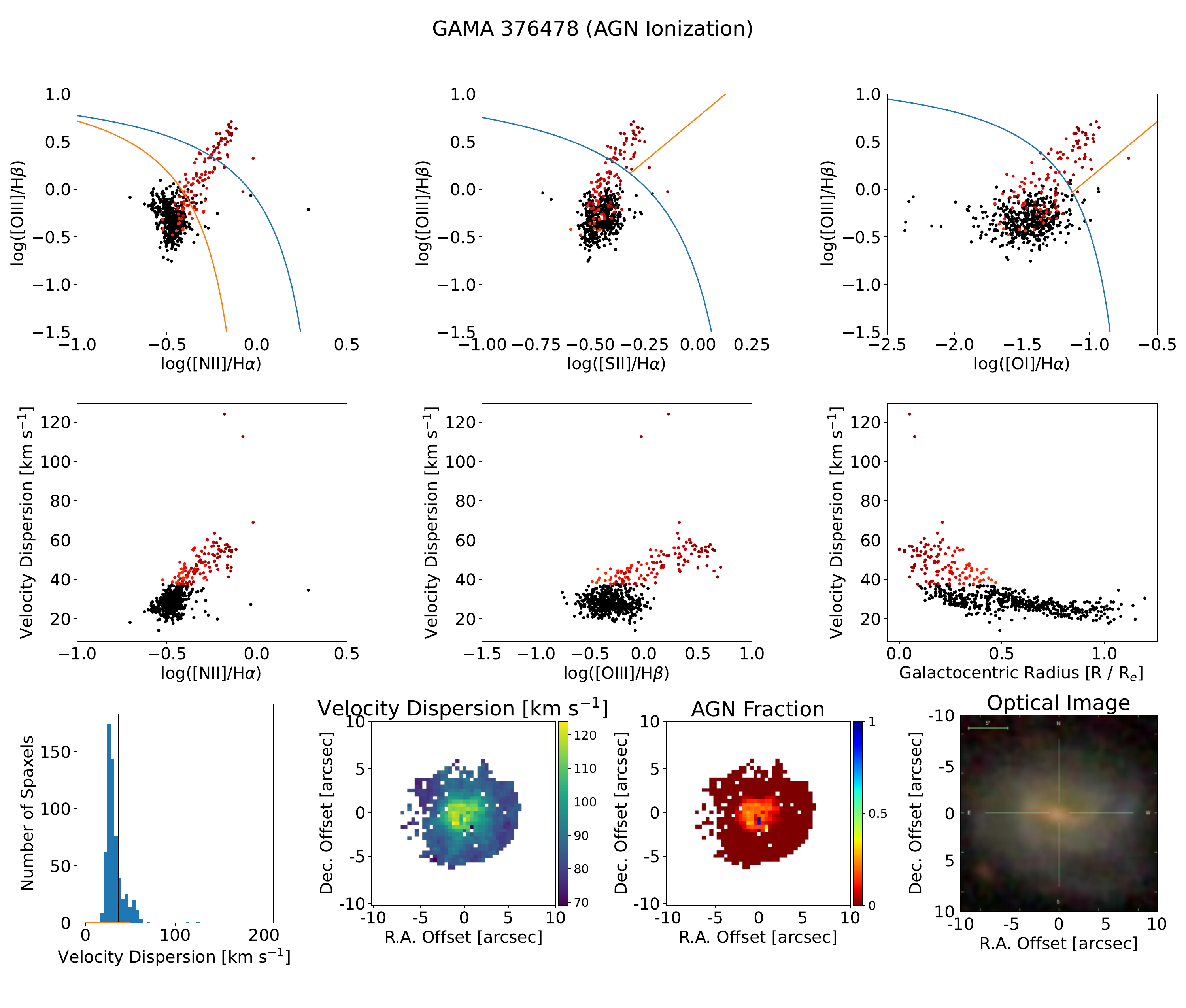}
   \caption{Emission line diagnostics and maps for GAMA 376478. The first row are the BPT/VO87 diagrams colored according to our new diagnostic, with black points being spaxels dominated by star-formation and colored points are spaxels with AGN ionization colored by increasing radius from the center of the galaxy. The middle row contains plots of log([\ion{N}{2}]/H$\alpha$), log([\ion{O}{3}]/H$\beta$), and effective radius versus velocity dispersion. The last row contains a histogram of the velocity dispersion with a vertical line denoting the two components in velocity dispersion space, a map of the velocity dispersion, a map of the AGN fraction which shows the relative contribution of AGN ionization on a per spaxel basis, and a three color image on the right matching the FOV of the SAMI data.} 
   \label{376478_panels}
 \end{figure}

Figures~\ref{376478_2d} \& \ref{376478_panels} shows a similar result as Figures~\ref{106717_2d} \& \ref{106717_panels}, but for GAMA 376478, which contains AGN excitation. Clear signatures of AGN ionization can be seen in the emission line diagnostics and spatial maps in Figure~\ref{376478_2d}, but more spaxels are detected that are traced further away from the nucleus of the galaxy with our revised method.

While our new diagnostic was tested with the data from the SAMI Galaxy Survey, the methods used are general enough to apply to any IFS data. As long as the spectral resolution is able to resolve roughly 30~km\,s$^{-1}$ in velocity dispersion to separate the non star-forming gas from the excited regions, the proposed method is viable. An increase of spatial resolution also helps to further separate the mixing sequences from one another, as shown in \citet{2019MNRAS.487.4153D}.

\section{Comparison with standard diagnostics}\label{compare}

Here we compare the results of our new classification scheme with those that use the classical BPT/VO87 diagrams without the additional use of kinematic or equivalent width information. For this work, we label a galaxy as star-forming, AGN, or shocked in emission line space only if there are a minimum of 10 spaxels or greater in each of the three BPT/VO87 diagrams in their respective regions after applying the same signal to noise cutoff as our revised method. This strict criterion ensures that we are not including low signal galaxies that could inflate the number of the AGN and shocked regions. Of the 3068 galaxies in the SAMI sample, 1996 of them remain after the signal to noise cut. From this remaining sample, using only BPT/VO87 diagnostics, we find that 233 galaxies contain clear signs of shocked gas, 50 host signatures of AGN ionization, and 33 show signatures of both shock and AGN ionization simultaneously. The remaining 1680 galaxies are identified as primarily hosting only ionization associated with \ion{H}{2} regions or star formation.

\begin{figure}[ht]
 \centering
   \includegraphics[width=.95\linewidth]{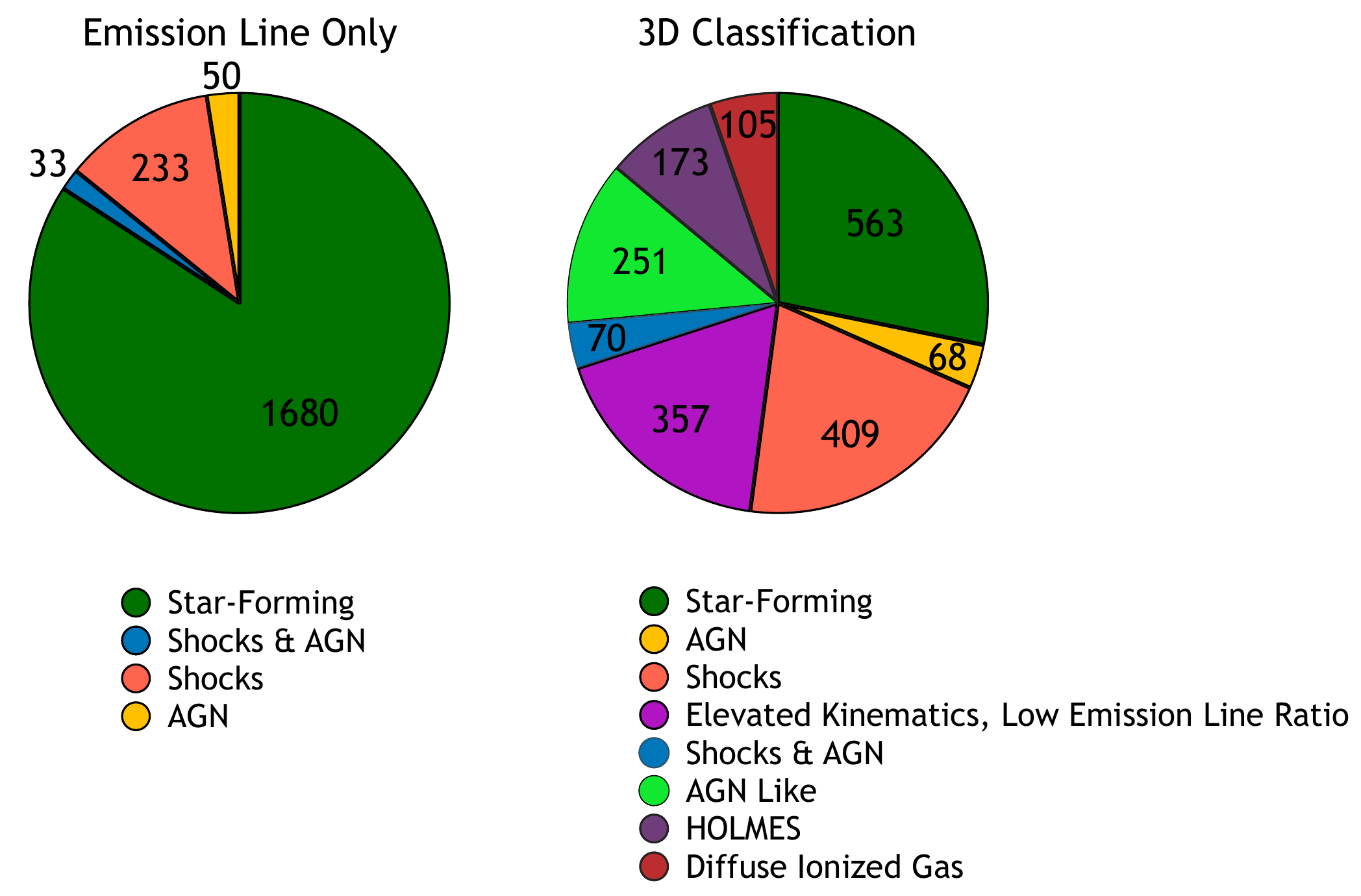}
   \caption{A comparison of ionization sources detected using just the BPT/VO87 diagrams (left) and our new method using emission line ratios, kinematics, and spatial information (right). Our 3D classification method is more sensitive at detecting ionization from non-star-forming sources than just emission line diagnostics alone.}
   \label{pie_chart}
 \end{figure}

For comparison, our new classification scheme detects 409 galaxies containing signatures of pure shock ionization, 68 containing only AGN ionization signatures, 70 galaxies containing shock and AGN signatures, 173 HOLMES contaminated galaxies, 105 galaxies with diffuse ionized gas signatures, 251 galaxies with AGN-like ionization, 563 pure star-forming galaxies, and 357 galaxies with elevated velocity dispersion and low emission line ratios. Figure~\ref{pie_chart} shows the results of our new classification scheme compared to BPT/VO87 emission line diagnostics alone.

For galaxies identified as hosting kinematically disturbed AGN or shocks, we also see an increase in the number of spaxels identified as being ionized by shocks or AGN on a galaxy by galaxy basis due to being able to trace the mixing sequence further into the star-forming region of the emission line diagnostic diagram. This can be most easily seen by comparing the emission line ratio only classification in Figures \ref{106717_2d} and \ref{376478_2d} with their new classification counterpart in the ionization fraction map in Figures \ref{106717_panels} and \ref{376478_panels} respectively. We will discuss here in more depth each of the categories of classification.

\subsection{Star Formation Dominated}
There are 563 galaxies identified as being dominated by ionization from \ion{H}{2} regions with no signs of AGN or shock excitation. These galaxies all have velocity disperisons of less than 50~km\,s$^{-1}$, which is consistent with star-formation dominated regions \citep{Rich2011,2015ApJS..221...28R,2014MNRAS.444.3961D}. The emission line ratios of these galaxies also lie below the maximum theoretical star-burst line, indicating a weaker source of ionization then is typically found with AGN or shocks. The combination of low velocity dispersions and low emission line ratios strongly suggests that these galaxies lack any significant contribution to their ionization from either widespread shocks or AGN, and are therefore classified as star formation dominated.

\subsection{Shocked Galaxies}
We identify 409 galaxies as hosting clear signatures of shock excitation, as opposed to 233 galaxies using emission line diagnostics only. Each of these galaxies has elevated velocity dispersions that are typical of turbulence caused by the mechanical nature of collisionally excited gas. These emission line diagnostics correlate with LINER-like ionization. A mixing sequence can be drawn for each of these galaxies by tracing the kinematically disturbed gas from low velocity dispersion, low emission line space to higher values of each. This allows us to follow the shock excitation from the most extreme values down to the the base of the star-forming sequence. Figure \ref{106717_panels} shows an example of this mixing sequence in more detail. By using this mixing sequence, we often trace the shock excitation below the standard classification lines as the sequence moves from high to low shock contribution.

\subsection{Separating HOLMES from AGN} \label{AGN_HOLMES}
Here we explore a further step used to ensure the purity of our AGN classification. HOLMES can present itself as an increased emission line ratio that mimics that of AGN or shocks \citep{1994_Binette,2008_Stasinska,2010_FernandesA,2010_FernandesB}. One reason for this is that an evolved stellar population that lacks young stars produces ionization that matches that of non-star-forming regions in emission line diagnostics. These hot low-mass evolved stars can have a significant contribution to the ionization of a galaxy, especially in galaxies with a lack of young stars such as ellipticals and early type spiral galaxies \citep{2018_Lacerda,2021_Sanchez,2020_Sanchez_review}. One way to account for this is to use the equivalent width of H$\alpha$ as a tracer of these regions. \citet{WHAN} has shown that spectra in which HOLMES dominates tend to have $\mathrm{EW}_{\mathrm{H}\alpha}$ $\leq$ 3 $\AA$.

\begin{figure}[ht]
 \centering
   \includegraphics[width=.95\linewidth]{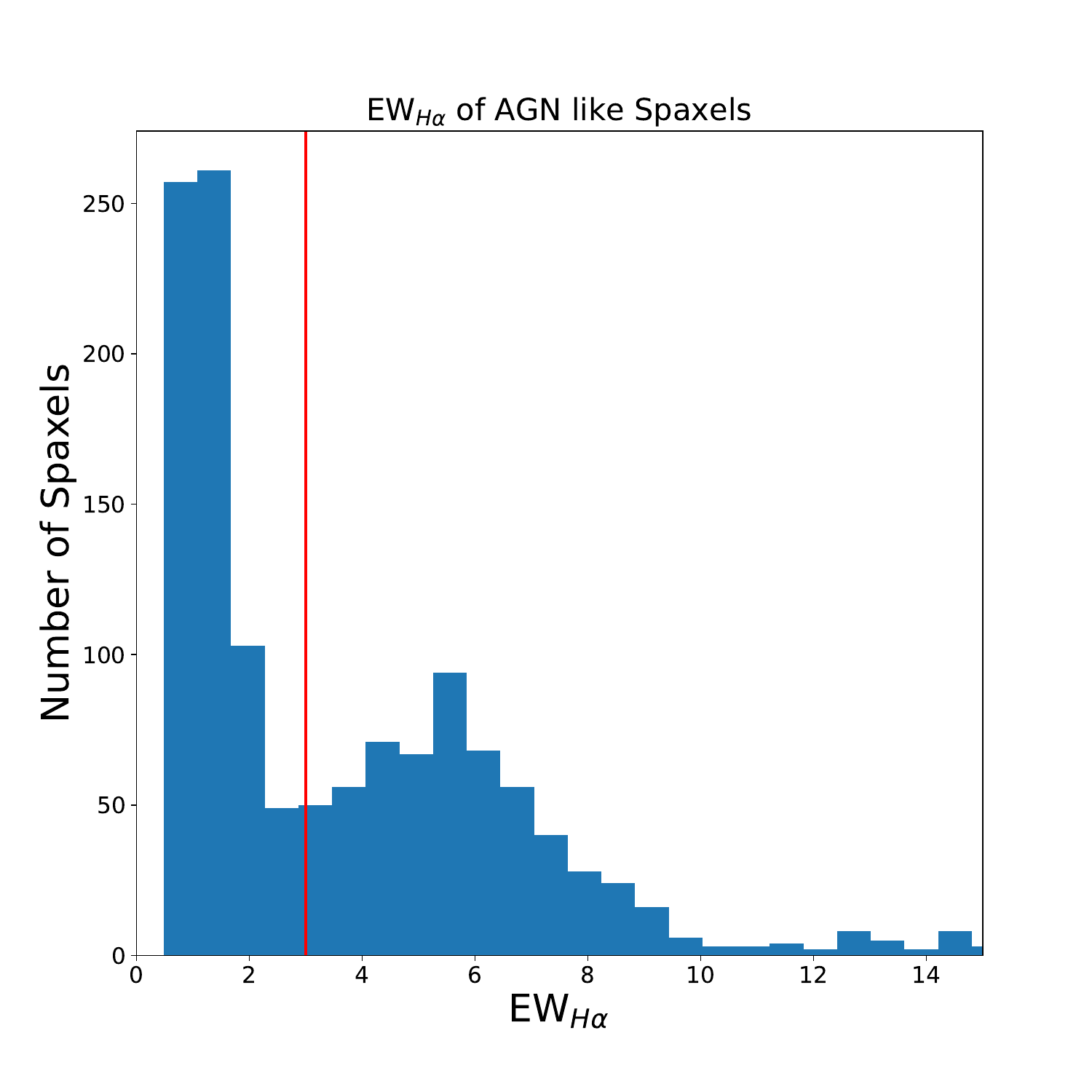}
   \caption{A histogram of the distribution of $\mathrm{EW}_{\mathrm{H}\alpha}$ for AGN-like spaxels. The red vertical line shows the 3 Angstrom separation used to distinguish between HOLMES and true AGN ionization.} 
   \label{EW_Histogram}
 \end{figure}

To this end, we perform an additional check on all of the galaxies identified as containing AGN-like ionization in our sample to investigate the contribution and possible contamination of HOLMES in these galaxies. When looking at the distribution of equivalent widths of H$\alpha$ in our AGN sample before accounting for HOLMES, we can see a clear bimodal distribution with a separation between two populations at an $\mathrm{EW}_{\mathrm{H}\alpha}$ of 3 Angstroms as shown in Figure \ref{EW_Histogram}. This distribution strongly suggests that a portion of our AGN galaxies show traces of HOLMES contamination at some level. Since it is possible that an AGN can present AGN-like $\mathrm{EW}_{\mathrm{H}\alpha}$ for the nuclear region before falling off to lower values radially, we do not apply a global cutoff to all spaxels of less than 3 angstroms. Instead, we look at the fraction of AGN-like spaxels that lie above and below the $\mathrm{EW}_{\mathrm{H}\alpha}$ cutoff on a per galaxy basis. The selection criteria used here requires that 70\% of a galaxy's spaxels that we identify as AGN ionized must have an $\mathrm{EW}_{\mathrm{H}\alpha}$ of less than 3 angstroms before being reclassified as HOLMES. This is done so that galaxies with weak AGN are not missed if only the central spaxels have a large $\mathrm{EW}_{\mathrm{H}\alpha}$ while spaxels further from the center are lower. Of the 311 AGN-like galaxies initially identified with our schema, 173 of these galaxies contain 70\% or more of their AGN spaxels showing HOLMES like equivalent widths that we reclassify. This brings our total number of AGN galaxies to 138. This combination of emission line ratios, gas kinematics, and $\mathrm{EW}_{\mathrm{H}\alpha}$ all help ensure that the galaxies that we label as AGN do indeed host a non-negligible fraction of AGN ionization.

\subsection{AGN Hosting Galaxies}\label{AGN}

After accounting for HOLMES ionization, our sample is left with a total of 138 AGN galaxies; 43 of which are identified from emission line diagnostics alone, 70 show elevated velocity dispersions and signatures of AGN and LINER ionization, and 25 show elevated velocity dispersions and AGN signatues, but no significant LINER ionization. We provide these different avenues for classification due to the different physical natures of each of them. The 43 emission line only galaxies are selected without using any kinematic information as a purely radiative AGN without shocks or outflows can still cause extreme ionization without mixing up the surrounding gas. This process would present itself as an increased emission line ratio while having not necessarily having increased velocity dispersion. We require agreement in all three of the BPT/VO87 diagrams in order to classify a galaxy as hosting an AGN due to the fact that galaxies that only have AGN-like ionization in two out of three emission line diagnostics look much different than those that have all three. We explore these two out of three galaxies in more detail in Section \ref{AGN_like}.

The 95 AGN selected with enhanced emission lines and velocity dispersions are likely indicative of AGN with some form of gas flow. These galaxies have AGN-like emission while also having velocity dispersions not associated with pure star-formation or a purely radiative AGN. 70 of these 95 galaxies also have significant amounts of LINER-like ionization detected as well, further supporting the possibility of shock induced AGN outflows. While many of these galaxies would classify as both shock and AGN ionized, we label them as AGN in the final classification as it is likely that the shocks are being driven by the AGN in the center of these galaxies.

\subsection{AGN-Like Galaxies} \label{AGN_like}

\begin{figure}[ht]
 \centering
   \includegraphics[width=.95\linewidth]{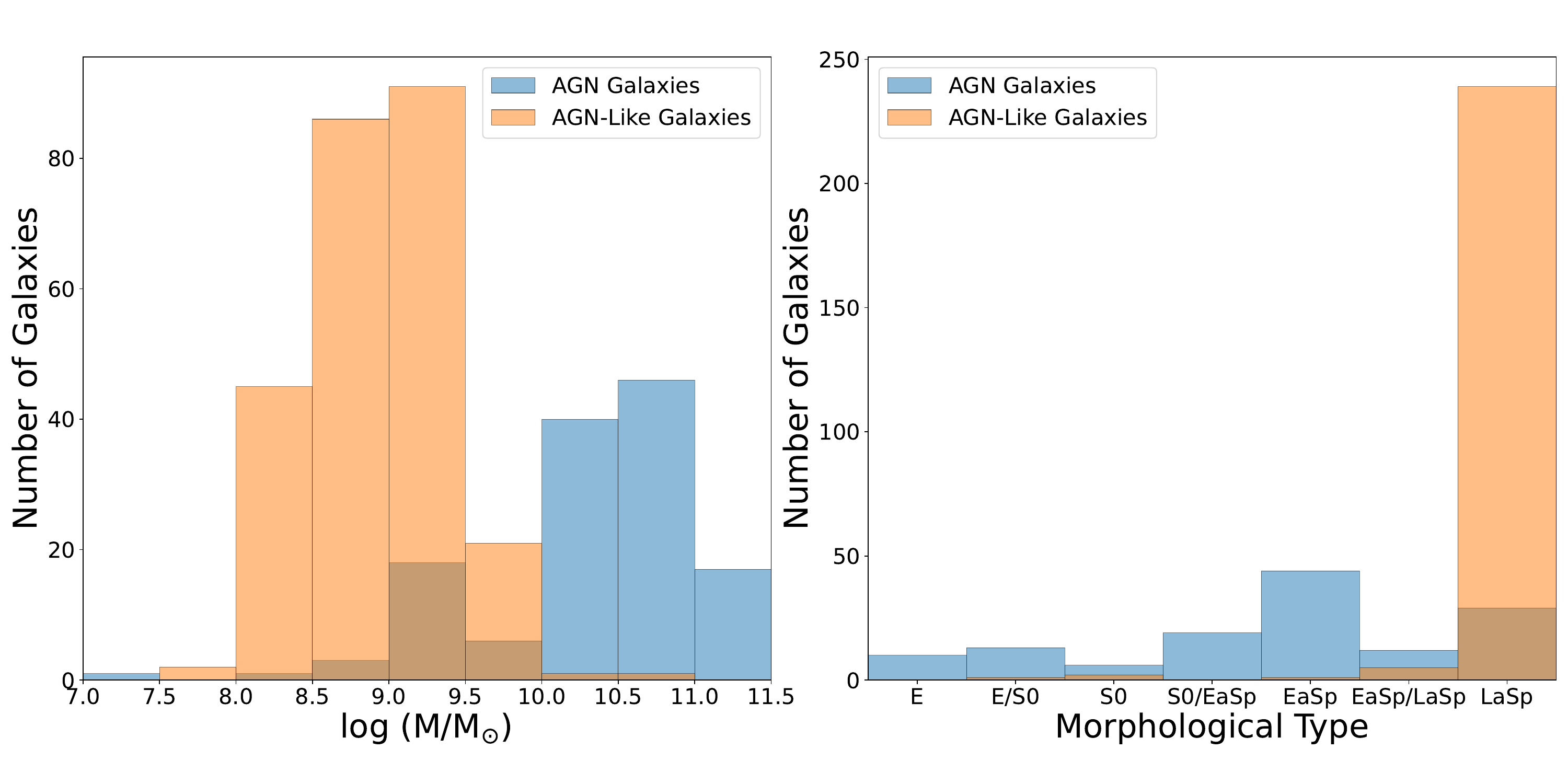}
   \caption{Histograms showing the distribution of galaxy mass and morphological type for galaxies identified as hosting AGN and AGN-like ionization.}
   \label{AGN_Hist}
 \end{figure}

\begin{figure}[ht]
 \centering
   \includegraphics[width=.95\linewidth]{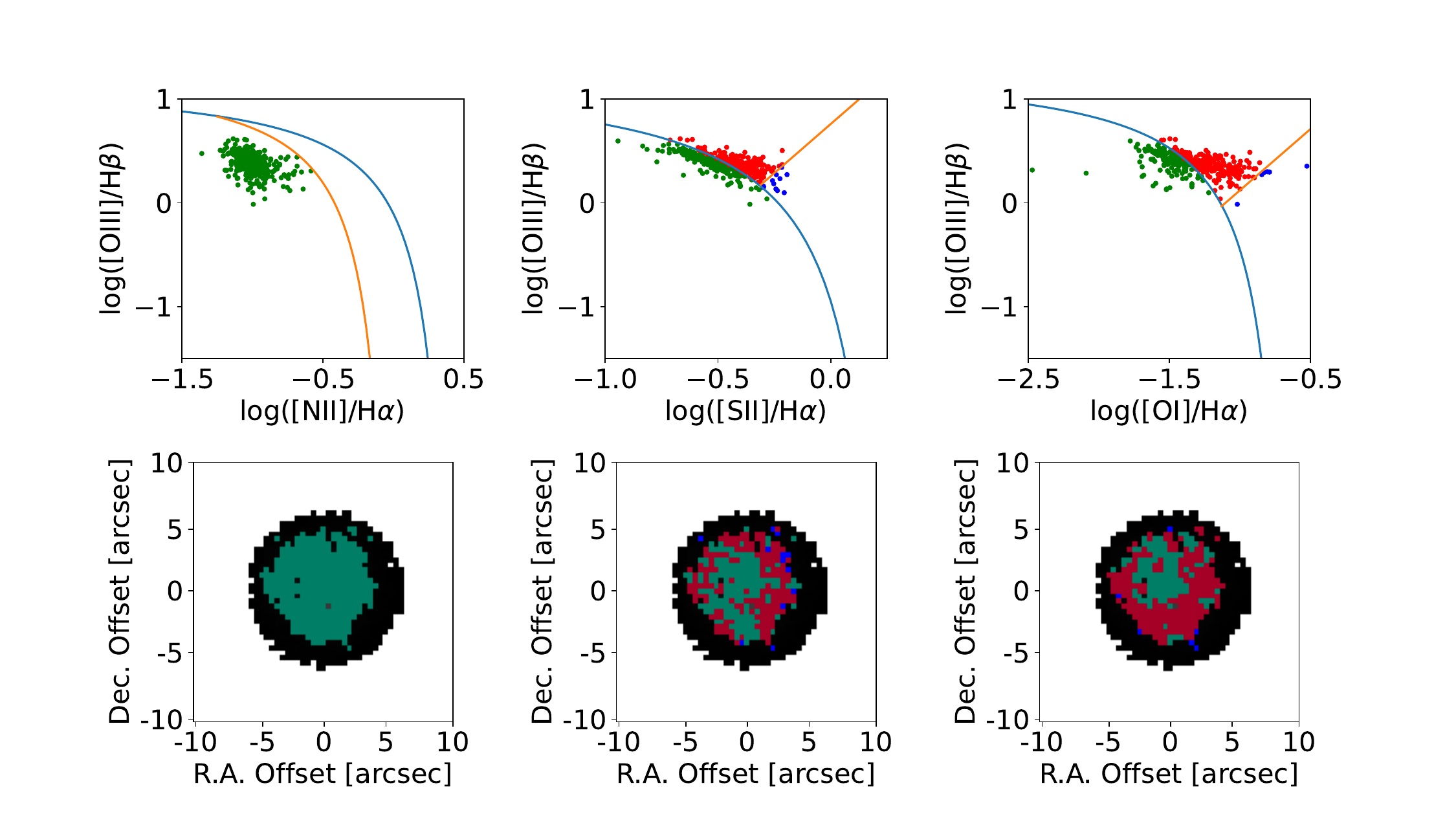}
   \caption{GAMA 273514 which contains AGN-like ionization. The top row shows BPT/VO87 emission line diagrams and the bottom row are spatial maps colored according to the diagnostic directly above it. There is no clear pattern in any of the spatial maps that suggest true ionization from a central AGN, despite the emission line diagnostics having a large number of AGN-like spaxels.}
   \label{AGN_like_figure}
 \end{figure}

We denote a special category of AGN-like galaxies that are separate from standard AGN. This category contains 251 galaxies and is characterized as having low velocity dispersions and exactly 2 out of 3 AGN-like emission line diagnostics. They also are not contaminated by HOLMES, as those cases have been moved into the proper category already. These cases stand out as being unlike a typical AGN galaxy in that there is no spatial coherence between the AGN-like spaxels within a galaxy. The mass of these galaxies also tends to be much lower than that of the elevated velocity dispersion or 3 out of 3 AGN sample as shown in figure \ref{AGN_Hist}. The morphological type is almost exclusively late spirals as well, with only 10 out of 251 galaxies not being a late spiral type. The location of spaxels on the emission line diagnostics is also unique in that they tend to clump together toward the low metallicity regions in the top left of each diagram, and extend nearly horizontally into the AGN region in the [\ion{S}{2}] and [\ion{O}{1}] diagrams, sometimes reaching into the LINER region as well. Figure \ref{AGN_like_figure} shows GAMA 273514, which contains this AGN-like ionization. There is no noticeable structure in the spatial map for any of the AGN-like spaxels, and the [\ion{N}{2}] diagram does not show any AGN-like spaxels at all. Nearly all of the galaxies in this population follow the same emission line diagnostic and spatial trends as GAMA 273514.

Despite the often times large number of AGN-like spaxels in these galaxies, given the combination of an on average low galaxy mass, single morphology type, no trace of any elevated velocity dispersion, and lack of spatial structure in the emission line maps, we do not classify these galaxies as true AGN hosts but rather as hosting AGN-like emission. While it is certainly possible that a number of these galaxies do indeed host an AGN, to determine if they are true AGN hosts would require a more in depth look at each galaxy individually, which is beyond the scope of this survey wide study.

\subsection{Elevated Velocity Dispersion, Low Ionization} \label{ElKiLoEm}

\begin{figure}[ht]
 \centering
   \includegraphics[width=.95\linewidth]{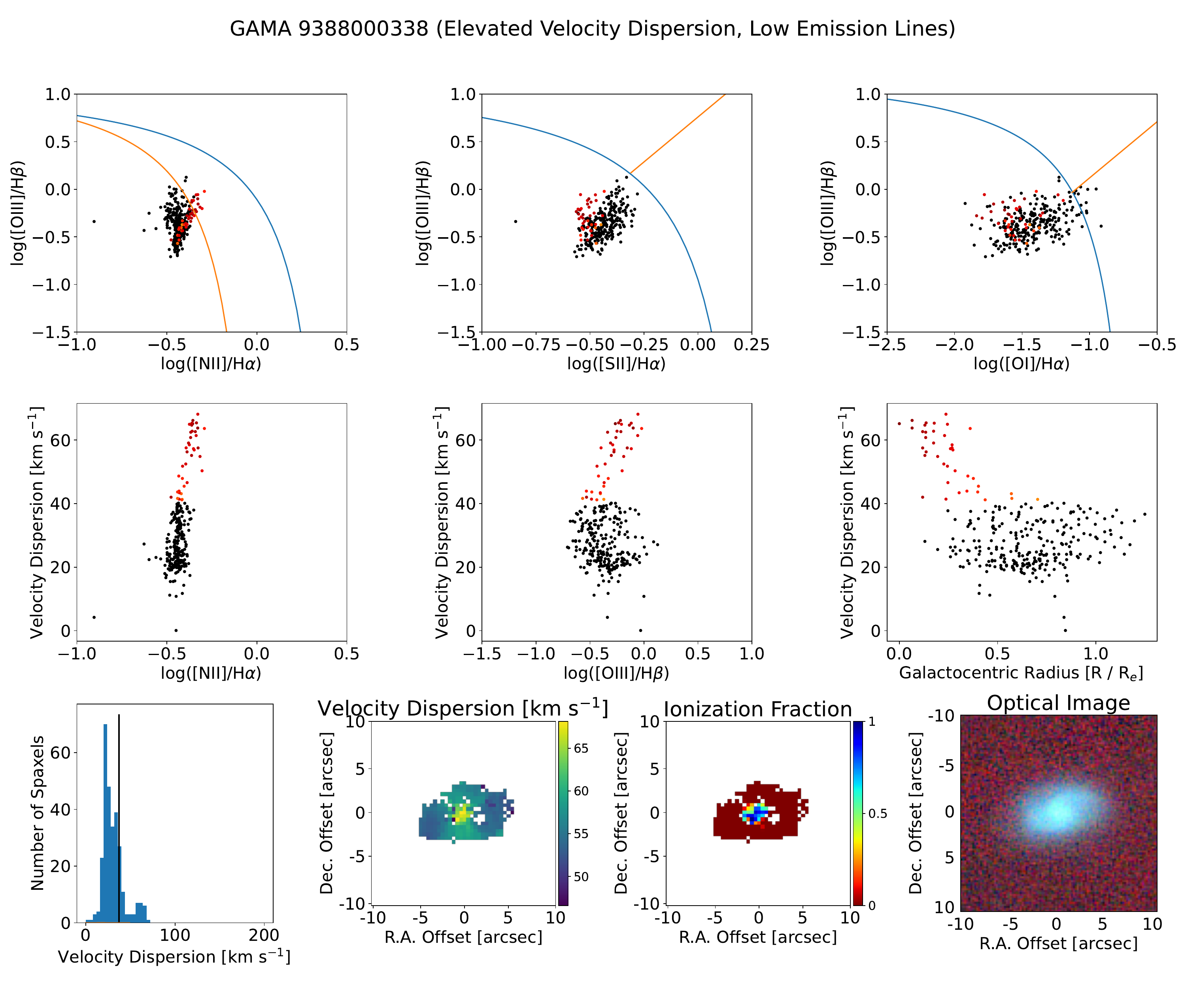}
   \caption{Emission line diagnostics and maps for GAMA 9388000338, a galaxy containing elevated velocity dispersion, low emission line spaxels. Each figure in the first row shows the emission line diagnostics colored according to where each spaxel fall in their respective diagram, green for star-forming, blue for composite or LINER, and red for AGN. Below each diagnostic is the corresponding spatial map showing where in the galaxy the points in the diagram above are located. The middle row contains plots of log([\ion{N}{2}]/H$\alpha$), log([\ion{O}{3}]/H$\beta$), and effective radius versus velocity dispersion with points being colored black if they lie below the velocity dispersion cutoff, and colored by increasing radius if they lie above the dispersion cutoff. The last row contains a histogram of the velocity dispersion with a vertical line denoting the two components in velocity dispersion space, a map of the velocity dispersion in the middle, and a three color image on the right matching the FOV of the SAMI data.} 
   \label{9388000338}
 \end{figure}

There are 357 galaxies we identify that have elevated velocity dispersion but few or no spaxels with high enough emission line ratios to be identified as shocks or AGN on their own. By using the kinematics, we can still identify a mixing sequence separate from the star-forming sequence, but are unable to trace it beyond the star-forming regions of BPT/VO87 line ratio space to identify whether the secondary ionization process is shocks or AGN. It is possible that a number of these galaxies are the result of beam-smearing artificially increasing the velocity dispersion. Because of the difficulty in separating between the effects of beam-smearing and real, physical processes within the galaxies for these cases, we label them simply as elevated velocity dispersion, low ionization and leave the final classification to the user based on their science goals. Figure~\ref{9388000338} shows an example of one of these galaxies.

\subsection{Diffuse Ionized Gas} \label{DIG}

A number of galaxies were found that have LINER-like ionization in emission line ratio diagnostics, but low velocity dispersions associated with pure star-formation. These galaxies are interesting in that while they have enough spaxels to be classified as non-star-forming in emission line space, they do not have kinematic signatures associated with shocks. These galaxies all share a similar trend of having regions of gas that extend into the LINER space in at least two of the BPT/VO87 diagrams while having velocity dispersions of $<$ 50~km\,s$^{-1}$ for the entire galaxy. The majority of their spaxels tend to lie along the maximum theoretical starburst line and don't typically extend much into higher emission line space. Their velocity dispersion profiles are all consistent with purely star-forming gas and show no signs of having two distinct tracks.

 \begin{figure}[ht]
 \centering
   \includegraphics[width=.95\linewidth]{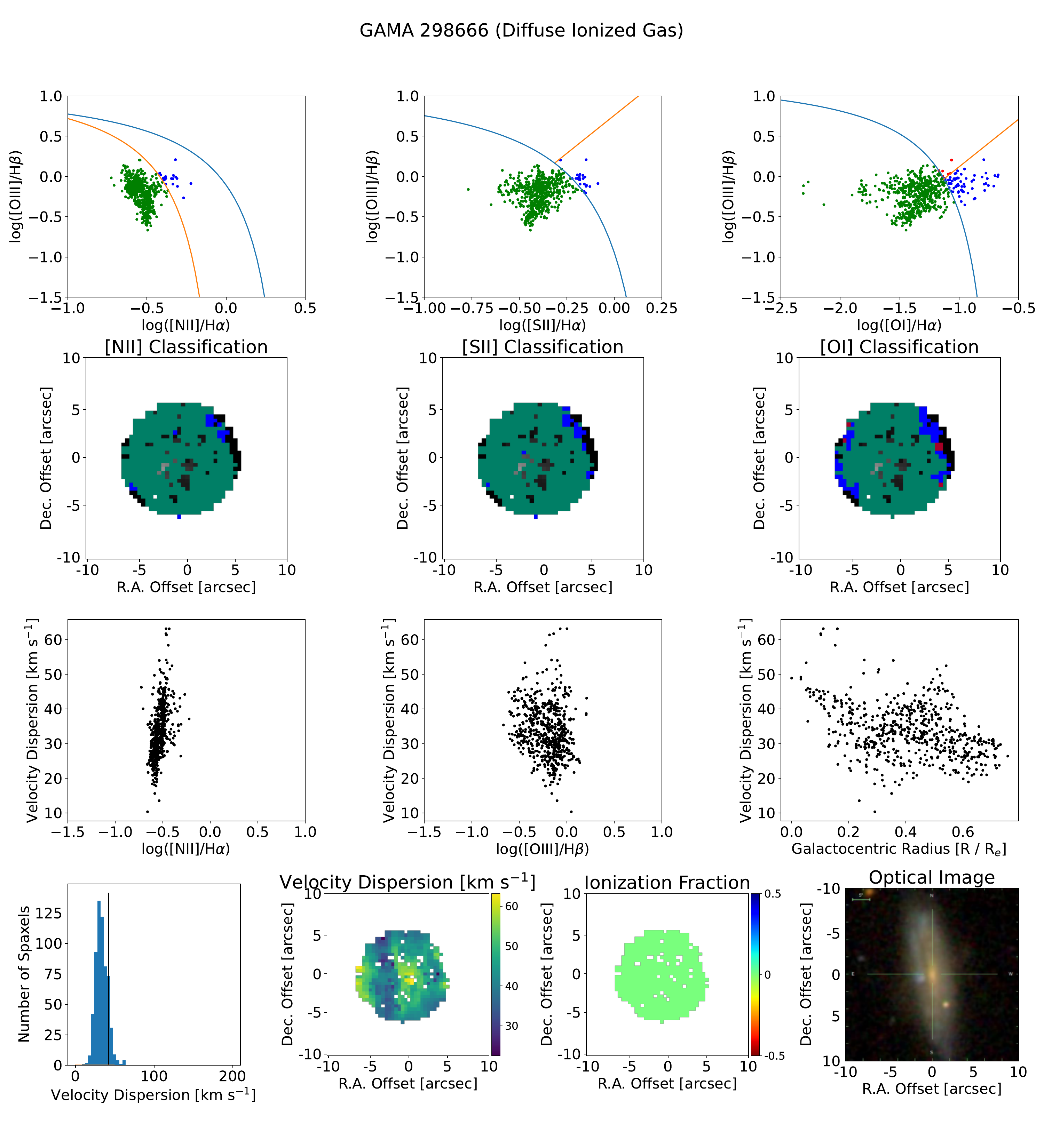}
   \caption{Emission line diagnostics and maps for GAMA 298666, a galaxy containing Diffuse Ionized Gas (DIG). Each figure in the first row shows the emission line diagnostics colored according to where each spaxel fall in their respective diagram, green for star-forming, blue for composite or LINER, and red for AGN. Below each diagnostic is the corresponding spatial map showing where in the galaxy the points in the diagram above are located. The middle row contains plots of log([\ion{N}{2}]/H$\alpha$), log([\ion{O}{3}]/H$\beta$), and effective radius versus velocity dispersion. As our method classifies all of the spaxels as being dominated by star-formation, all of the points here are labeled black. The last row contains a histogram of the velocity dispersion with a vertical line denoting the two components in velocity dispersion space, a map of the velocity dispersion in the middle, and a three color image on the right matching the FOV of the SAMI data.} 
   \label{298666}
 \end{figure}

For the 105 galaxies with LINER-like ionization and low velocity dispersions, one possible explanation is that they contain diffuse ionized gas, which would have elevated line ratios while having a lower velocity dispersion \citep{DIG73,DIG_PHANGS,DIG_300}. Traditionally, diffuse ionized gas is primarily detected in edge on spiral galaxies at low redshifts \citep{DIG1994,DIG2003}. The nature of diffuse ionized gas would align with the spread out nature of the elevated emission lines with low velocity dispersion. These elevated emission line ratios could also be caused by ionization from post-Asymptotic Giant Branch (AGB) stars. The radiation released from these stars can ionize the surrounding medium without mixing up the gas \citep{AGB1994,AGB2008,AGB2011}. Future in-depth analysis of these galaxies is needed to confirm the physical processes at play; here we simply classify them as diffuse ionized gas.

\section{Ionization Maps}\label{ionization_maps}

\begin{figure}[ht]
 \centering
   \includegraphics[width=.95\linewidth]{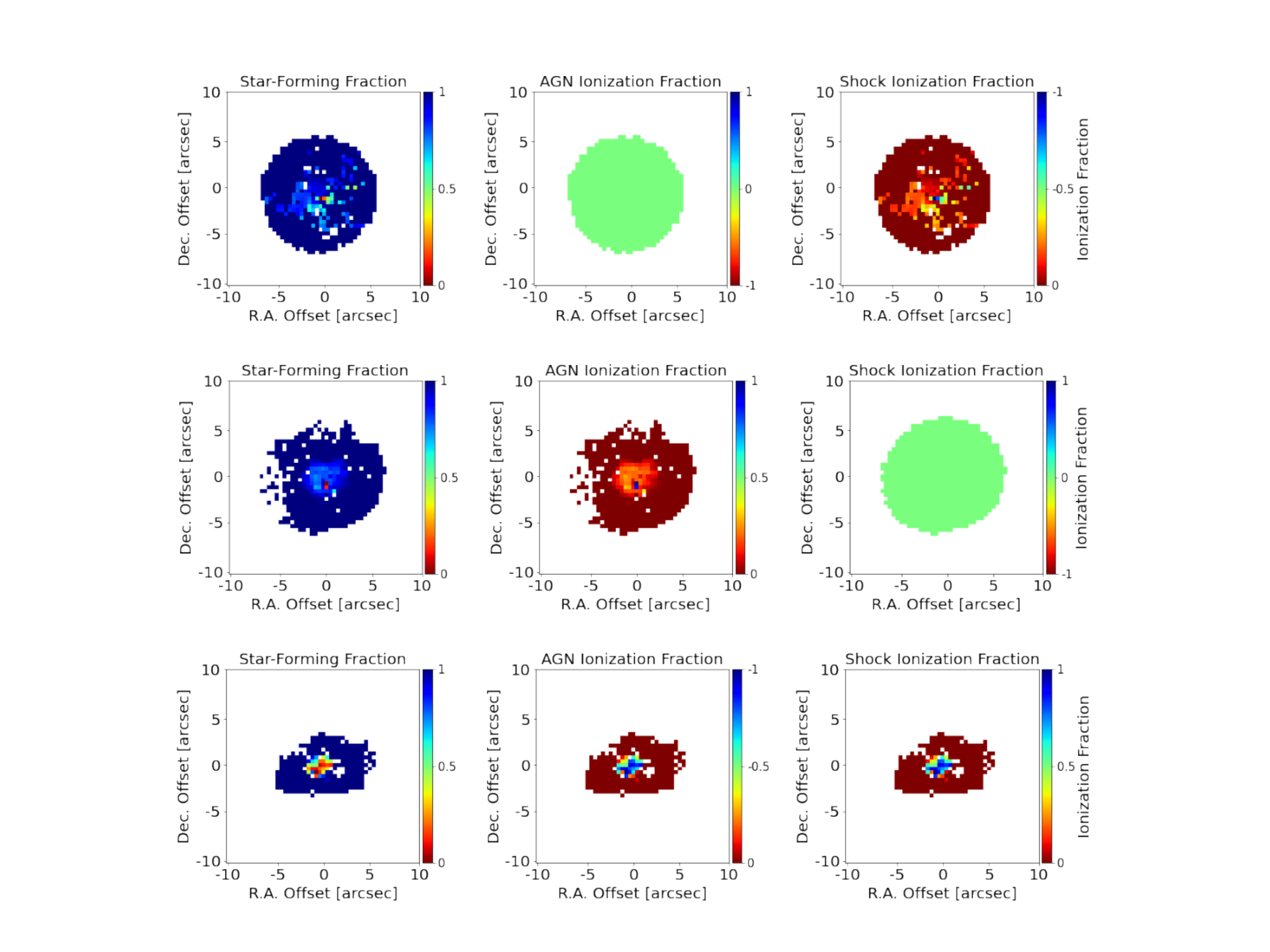}
   \caption{Ionization map examples for GAMA 106717, 376478, and 9388000338 respectively. The relative contributions of each driving power source is shown as a fraction of the total amount of ionization. For galaxies with either AGN or shock driven ionization, the corresponding layers are mapped according to the mixing sequence identified by our method. As our method can not separate ionization sources within a single galaxy, one layer will have no contribution to the total ionization. For galaxies where non star-forming ionization is detected but we can not definitively determine between shocks, AGN, and beam-smearing, a map of negative values is placed in both layers to denote this uncertainty.} 
   \label{shock_map}
 \end{figure}

As part of our analysis we have created ionization map data products that show the fractional contribution of each possible ionization source on a per spaxel basis for those galaxies that have elevated velocity dispersions. These values are calculated separately for each galaxy and are not global. The data products are image files consisting of three layers; star formation, AGN, and shock contribution. Each layer shows the spatial map of fractional contribution to the emission line fluxes by each ionization source. As we are unable to separate both AGN and shocks within a single galaxy, only the ionization source detected will have a non-zero map. The ionization amount is determined by weighing a spaxel's distance along the mixing sequence starting at the base of the star-forming track by the flux in H$\alpha$ for each component and summing the totals together. The mixing sequence is created by plotting a best fit line of the points that lie above the velocity dispersion threshold in 3D space from the lowest to highest values in velocity dispersion. This effectively traces the path from the base of the star-formation sequence up into the high velocity dispersion, non-star-forming regions. For the galaxies that host unidentified non-star-forming ionization, we list their non-star-forming component as negative values in both the AGN and shock maps to denote the uncertainty in what the major driver of this ionization is. An example of three galaxies with shock, AGN, and elevated velocity dispersion, low emission line ionization are shown in Figure \ref{shock_map}. Along with these ionization maps, we will release a catalog denoting which sources of ionization are detected in each galaxy. These data products will be released for public use alongside other SAMI Galaxy Survey Value-Added Products through the data central database \footnote{https://datacentral.org.au/)}.

\section{Conclusion} \label{conclusion}

We have investigated a new way of classifying the excitation sources of warm ionized gas in a galaxy using a new multi-dimensional diagnostic diagram. Our classification method takes advantage of the multiple levels of data that can be obtained by using integral field spectroscopy. By combining emission line ratios, gas kinematics, and spatial information, we determine the primary excitation source for individual spaxels of a galaxy. Our new method was tested using the SAMI Galaxy Survey, but is general enough to be used with any IFS data set.

Out of the 1996 galaxies used in our sample, our new method finds 409 galaxies with gas ionized by shocks, and 138 galaxies with gas ionized by a central AGN as compared to 233 and 88 respectively using emission lines diagnostics alone. We also identify and label sources of ionization not accounted for in standard emission line diagnostics such as from diffuse ionized gas and hot, low mass evolved stars. The combination of gas kinematics and emission line ratios enable better detection of low powered, non star-forming ionization sources. Coupled together with spatial information on a per spaxel basis, we are able to determine a star-forming to AGN or shock mixing sequence for those galaxies that are not dominated by pure star-formation.

We also classify a total of 356 galaxies as hosting a spatially incoherent ionization of both high emission line ratios and low velocity dispersions not typically associated with either pure star-formation, AGN, or shock ionization. 105 of these galaxies show LINER-like emission and 251 show AGN-like emission which we label as diffuse ionized gas and AGN-like respectively. We find an additional 173 galaxies that show increased emission line ratios and kinematics, but low $\mathrm{EW}_{\mathrm{H}\alpha}$ consistant with ionization from hot low-mass evolved stars. Finally, we find 357 galaxies containing non-star-forming like velocity dispersions that we are unable to distinguish between either low luminosity AGN, shocks, or beam-smearing. These galaxies are characterized as containing elevated kinematics and low emission line ratios.

In addition to the results presented in this paper, we will be releasing shock maps for all of the galaxies contained within SAMI DR3. These data products will show the fractional contribution of ionization on a per spaxel basis as well as the associated ionization source. The finalized shock maps will be available to download through \url{https://datacentral.org.au/}.


The authors wish to pay respect to the Gamilaraay/Kamilaroi language group Elders—past, present, and future—of the traditional lands on which the Siding Spring Observatory stands. VDJ and AMM acknowledge support from the National Science Foundation under grant number 2009416. JvdS acknowledges support of an Australian Research Council Discovery Early Career Research Award (project number DE200100461) funded by the Australian Government. Parts of this research were supported by the Australian Research Council Centre of Excellence for All Sky Astrophysics in 3 Dimensions (ASTRO 3D), through project number CE170100013. The SAMI Galaxy Survey is based on observations made at the Anglo-Australian Telescope. The Sydney-AAO Multi-object Integral field spectrograph (SAMI) was developed jointly by the University of Sydney and the Australian Astronomical Observatory. The SAMI input catalogue is based on data taken from the Sloan Digital Sky Survey, the GAMA Survey and the VST ATLAS Survey. The SAMI Galaxy Survey is supported by the Australian Research Council Centre of Excellence for All Sky Astrophysics in 3 Dimensions (ASTRO 3D), through project number CE170100013, the Australian Research Council Centre of Excellence for All-sky Astrophysics (CAASTRO), through project number CE110001020, and other participating institutions. The SAMI Galaxy Survey website is \url{http://sami-survey.org/}.

This paper includes data that has been provided by AAO Data Central (datacentral.org.au).

\section{Appendix: 3D Diagrams}

Here we include a number of three dimensional diagnostic diagrams for each of the galaxies discussed in this paper. Interactive versions of these figures can be found in the online version of this journal.

\begin{figure}[ht]
 \centering
   \includegraphics[width=.95\linewidth]{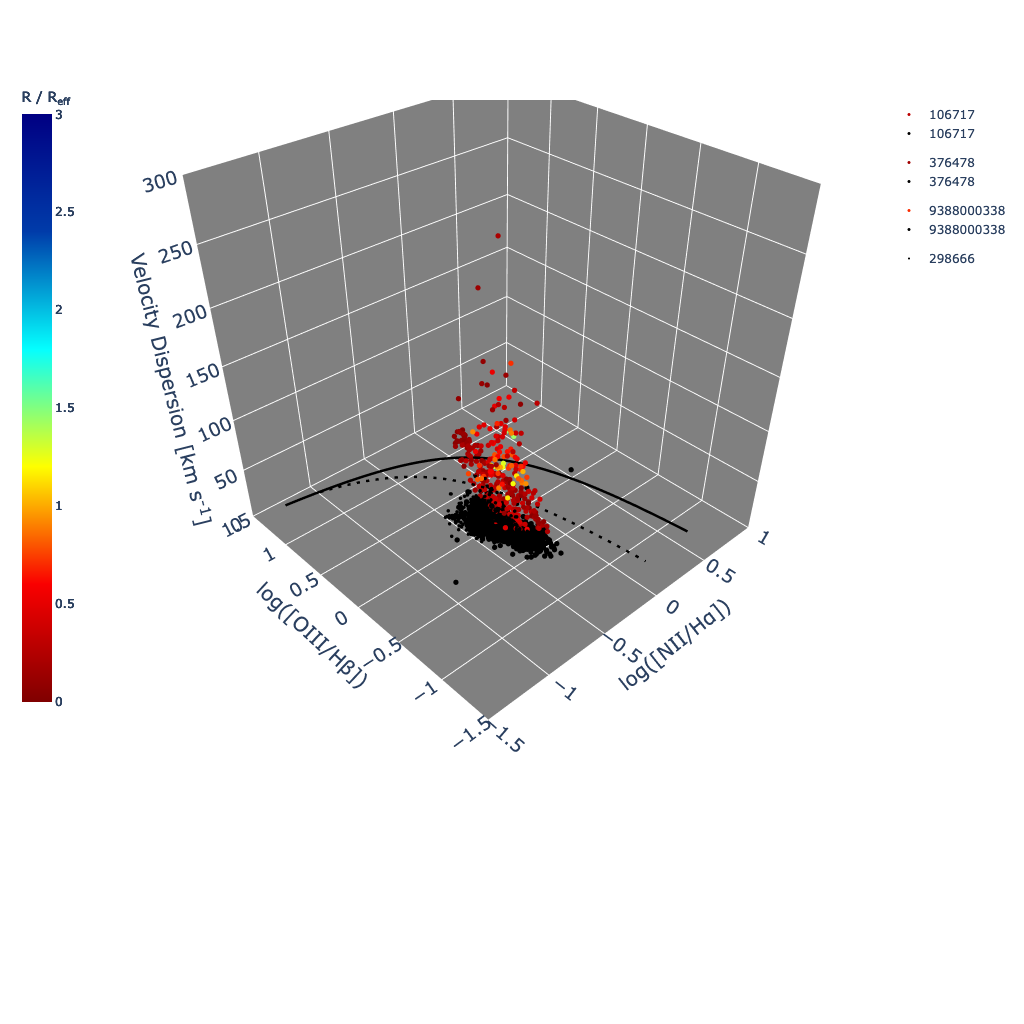}
   \caption{A three dimensional view of our new emission line diagnostic using the log([\ion{N}{2}]/H$\alpha$) emission line pair for GAMA 106717, 376478, 9388000338, and 298666. The online interactive figure allows for the plot to be rotated, zoomed in and out, and for each galaxy to be turned on and off in the figure by selecting it from the legend.} 
 \end{figure}
 
 \begin{figure}[ht]
 \centering
   \includegraphics[width=.95\linewidth]{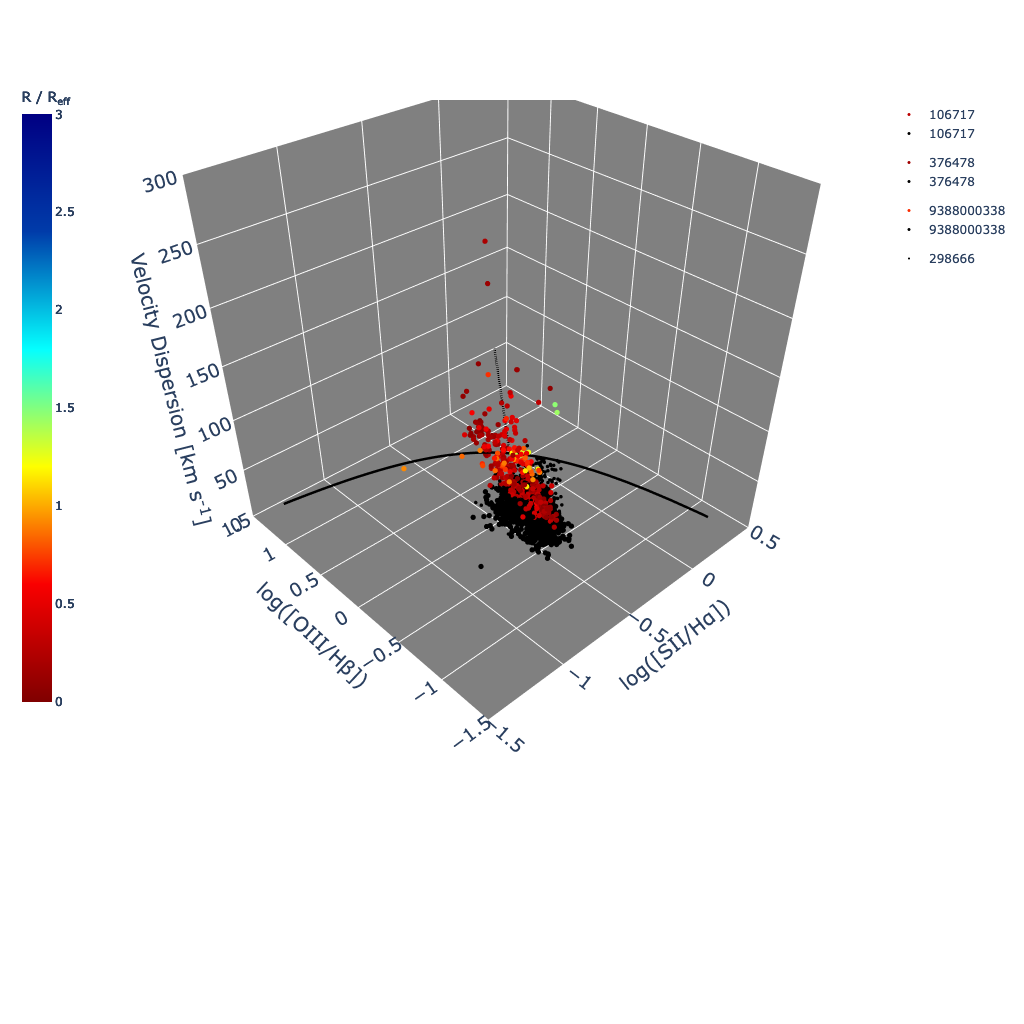}
   \caption{A three dimensional view of our new emission line diagnostic using the log([\ion{S}{2}]/H$\alpha$) emission line pair for GAMA 106717, 376478, 9388000338, and 298666. The online interactive figure allows for the plot to be rotated, zoomed in and out, and for each galaxy to be turned on and off in the figure by selecting it from the legend.}
 \end{figure}

\begin{figure}[ht]
 \centering
   \includegraphics[width=.95\linewidth]{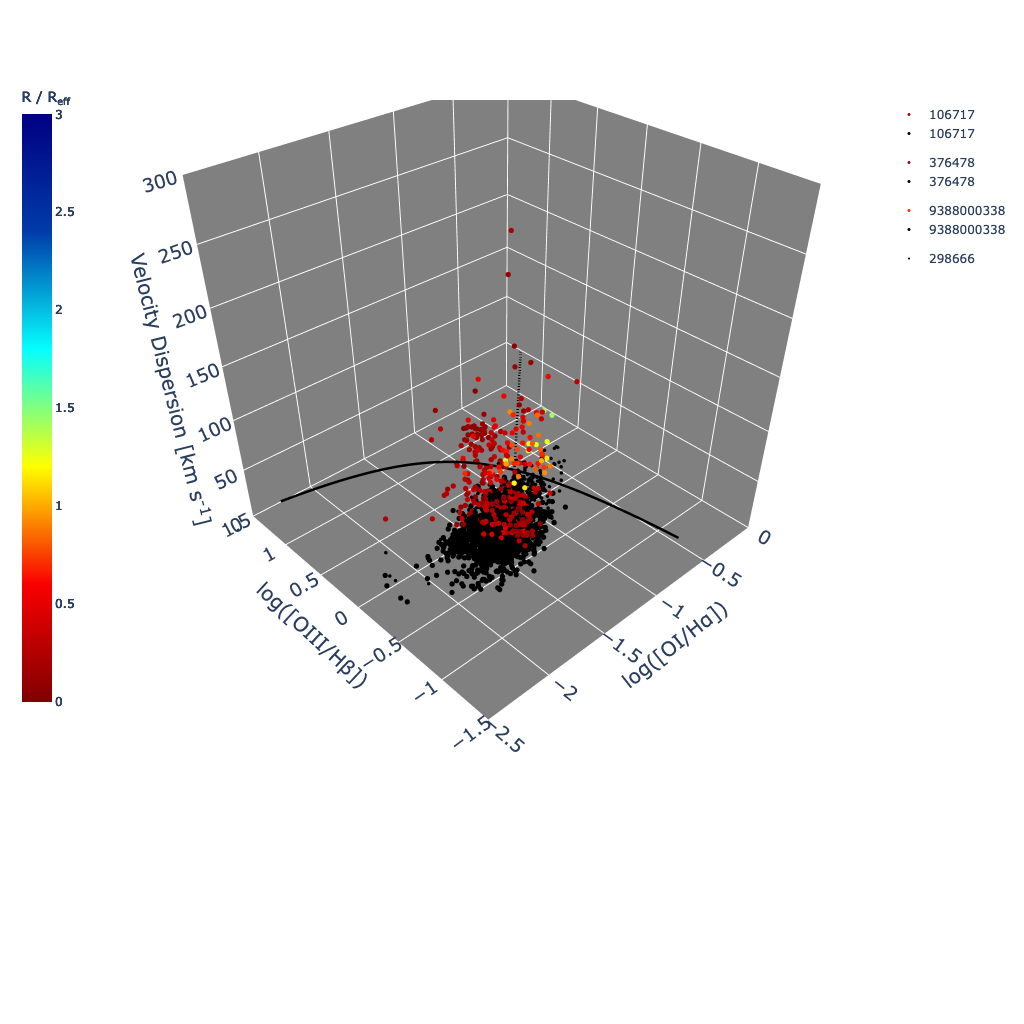}
   \caption{A three dimensional view of our new emission line diagnostic using the log([\ion{O}{1}]/H$\alpha$) emission line pair for GAMA 106717, 376478, 9388000338, and 298666. The online interactive figure allows for the plot to be rotated, zoomed in and out, and for each galaxy to be turned on and off in the figure by selecting it from the legend.}
 \end{figure}

\bibliography{References.bib}

\end{document}